\documentclass [10pt, twocolumn, conference] {IEEEtran}
\usepackage{amssymb}
\usepackage[english]{babel}
\usepackage[autostyle]{csquotes}
\usepackage{graphicx}
\usepackage[utf8]{inputenc}
\usepackage{textgreek}
\usepackage{booktabs} 
\usepackage{array} 
\usepackage[none]{hyphenat}
\let\ket\relax
\usepackage[braket, qm]{qcircuit}
\usepackage{physics}
\usepackage{dblfloatfix}
\usepackage{url}
\usepackage{multirow}
\usepackage{epstopdf}
\usepackage{setspace}
\usepackage{caption}
\usepackage{booktabs,fixltx2e}
\usepackage[flushleft]{threeparttable}
\usepackage{hyperref} 
\usepackage{algorithm,algpseudocode}
\usepackage{enumitem}
\usepackage{amsmath}
\usepackage{comment}
\usepackage{subcaption}
\usepackage{caption}
\usepackage{balance}
\usepackage{siunitx}
\usepackage{diagbox}
\graphicspath {{./images}}
\algrenewcomment[1]{\bgroup\hfill//~#1\egroup}
\usepackage{color}
\usepackage{romannum}
\usepackage{color,soul}
\usepackage{algorithm}
\usepackage{algpseudocode}
\usepackage[braket, qm]{qcircuit}

\begin{document}
\sloppy

\title{Pulse-Level Simulation of Crosstalk Attacks on Superconducting Quantum Hardware\textsuperscript{†}}
\author{Syed Emad Uddin Shubha, Tasnuva Farheen\\Division of Computer Science, Louisiana State University}

\maketitle

\let\thefootnote\relax\footnotetext{\textsuperscript{†}This paper has been accepted to the Security, Privacy, and Resilience Workshop at IEEE Quantum Week (QCE 2025) and will appear in the workshop proceedings.}

\begin{abstract}
Hardware crosstalk in multi-tenant superconducting quantum computers poses a severe security threat, allowing adversaries to induce targeted errors across tenant boundaries by injecting carefully engineered pulses. We present a simulation-based study of active crosstalk attacks at the pulse level, analyzing how adversarial control of pulse timing, shape, amplitude, and coupling can disrupt a victim's computation. Our framework models the time-dependent dynamics of a three-qubit system in the rotating frame, capturing both always-on couplings and injected drive pulses. We examine two attack strategies: attacker-first (pulse before victim operation) and victim-first (pulse after), and systematically identify the pulse and coupling configurations that cause the largest logical errors. Protocol-level experiments on quantum coin flip and XOR classification circuits show that some protocols are highly vulnerable to these attacks, while others remain robust. Based on these findings, we discuss practical methods for detection and mitigation to improve security in quantum cloud platforms.
\end{abstract}

\vspace{1em}
\begin{IEEEkeywords}
Quantum computing, Crosstalk, Side-channel attack, Quantum hardware security, Pulse-level control
\end{IEEEkeywords}


\section{Introduction}
The rapid growth of multi-tenant, cloud-accessible quantum computing has introduced new challenges in reliability and security. In superconducting architectures, quantum computation depends on the ability to couple qubits to implement entangling gates and scalable logic. However, this same coupling creates vulnerabilities. Among these, \emph{hardware crosstalk}, where control pulses or interactions intended for one qubit unintentionally influence others, remains a subtle but critical risk. Crosstalk arises from inevitable hardware design choices, such as capacitive or inductive coupling, and often manifests as persistent, always-on interactions that shift a qubit's energy levels based on the state of its neighbors~\cite{mundada2019suppression}. In addition to creating unwanted ZZ interactions, these parasitic couplings can also introduce off-diagonal ZX and YX terms that are particularly effective channels for inducing state rotations, as we will demonstrate.

Traditionally treated as a calibration or fidelity problem~\cite{ash2020analysis, rudinger2021experimental, zhao2022quantum, li2022pulse}, crosstalk is now recognized as a potential vector for adversarial attacks in real-world shared hardware. While hardware engineers view crosstalk as a fidelity challenge, recent studies have demonstrated that it can be exploited as a physical-layer side channel or an \emph{active attack mechanism} in multi-user quantum devices~\cite{ash2020analysis, maurya2024understanding, choudhury2024crosstalk}. By injecting tailored pulses into their assigned qubits, an attacker can induce targeted errors in a victim’s computation, such as flipping the phase of a target qubit to alter a superposition state~\cite{xu2024jailbreaking}. Critically, such attacks can degrade computational integrity, bias outcomes, or introduce errors without requiring direct access to the victim’s circuit, making them a serious threat to the security of shared quantum resources.

Despite this, prior research on active attacks has largely focused on two areas. The first involves an adversary who directly modifies the pulse-level definitions within a victim's circuit, exploiting software and interface vulnerabilities to alter pulse timing, frequency, or waveform~\cite{xu2024jailbreaking}. The second uses crosstalk primarily as a passive side-channel to learn about a victim's circuit structure~\cite{ash2020analysis, maurya2024understanding, choudhury2024crosstalk}. While the idea of active crosstalk attacks has been noted in prior literature, these works often emphasize circuit-level or conceptual models. 
In contrast, our work focuses on the threat model where the attacker applies adversarial pulses to \textit{their own} qubits without modifying the victim’s code, intentionally inducing errors on a neighboring victim qubit through the hardware's inherent physical coupling, and analyzes its effects through detailed time-dependent Hamiltonian simulations. Such attacks are particularly insidious as they operate below the standard gate-level abstraction and leverage a purely physical phenomenon, potentially evading detection by conventional error mitigation schemes.

In summary, this paper presents a detailed, Hamiltonian-based framework to simulate and analyze active crosstalk attacks. We first define our threat model and simulation methodology, then systematically identify the most potent adversarial configurations by varying pulse parameters and coupling types. We then evaluate their impact through case studies on two distinct quantum protocols, revealing that vulnerability is highly protocol-dependent. Finally, based on these simulation insights, we conclude with a discussion of practical mitigation strategies.

\section{Threat Model}\label{sec:threat_model}
\begin{figure}[!htb]
    \centering
    \includegraphics[width=0.7\columnwidth]{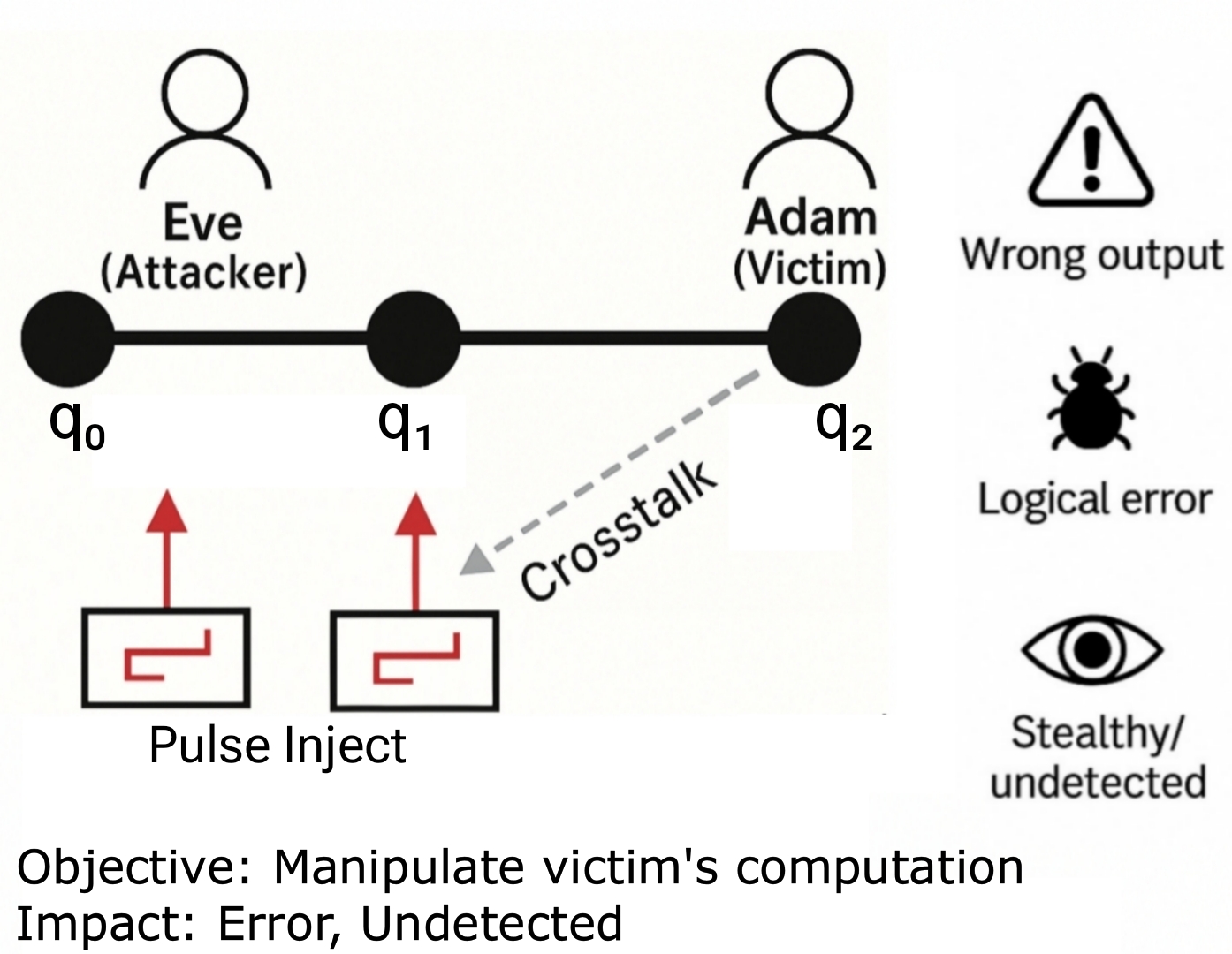} 
    \vspace{-5pt}
    \caption{\small Illustration of a crosstalk attack. The attacker (Eve) injects microwave pulses into $q_0$ and $q_1$, creating crosstalk that perturbs the victim's qubit ($q_2$), used by Adam.}
    \vspace{-9pt}
    \label{fig:crosstalk_attack}
\end{figure}

In this work, we consider a multi-tenant quantum cloud setting where two users, Eve (the attacker) and Adam (the victim), share a device with three linearly connected qubits. Eve has pulse-level control over $q_0$ and $q_1$, which are physically adjacent to Adam's qubit, $q_2$. Eve's goal is to exploit the device's inherent crosstalk pathways to corrupt Adam's computation, as illustrated in Figure~\ref{fig:crosstalk_attack}.

We make the following assumptions about the threat model:
\begin{itemize}
    \item Eve has complete, pulse-level control over the parameters (shape, amplitude, frequency, timing) for $q_0$ and $q_1$ but cannot directly access $q_2$ or Adam's circuit.
    \item Adam uses $q_2$ through standard gate-based instructions and is unaware of any attack, attributing any resulting errors to normal hardware noise or operational drift.
\end{itemize}

Under this model, we compare two attack strategies: \textit{attacker-first}~(adversarial pulses precede victim’s operation) and \textit{victim-first}~(adversarial pulses follow). We use Hamiltonian-based simulation to study active crosstalk attacks at the pulse level. Our approach models both always-on coupling and adversarial drive pulses in a three-qubit chain. We systematically scan pulse shapes and coupling types to identify which configurations most effectively induce errors and evaluate protocol-level impact on tasks such as a quantum coin flip and an XOR classification~\cite{grossu2021single}. Our results demonstrate that certain quantum protocols are highly vulnerable to these pulse-level attacks, while others remain robust, highlighting need for protocol-aware security analysis.

\section{Attack Model and Simulation Framework}
While Section~\ref{sec:threat_model} defines the overall threat model and assumptions, this section focuses on the specific attack strategies and system-level pulse modeling used to implement and analyze the threat.
To investigate pulse-level crosstalk attacks, we developed a framework that combines strategic adversarial timing with a detailed physical model of the quantum system. This section outlines the system Hamiltonian, the adversarial strategies, the specific pulse parameters under attacker control, and the numerical methods used for simulation and analysis.

\subsection{System Dynamics and Hamiltonian}
We model the three-qubit system's evolution using a time-dependent Hamiltonian in the \textbf{rotating frame}. This standard technique removes the fast-oscillating terms associated with qubit frequencies, allowing the simulation to focus efficiently on the slower dynamics induced by control pulses and parasitic couplings, which are central to the attack~\cite{wei2022hamiltonian, balewski2025first}.

The total Hamiltonian is composed of a static coupling term and a time-dependent drive term:
$$
\small H(t) = H_{\text{coupling}} + H_{\text{drive}}(t)
$$

\noindent\textbf{Crosstalk Hamiltonian~($H_{\text{coupling}}$):} This static term models the persistent, always-on parasitic interactions between physically adjacent qubits, a common feature in superconducting hardware. We define it as:
$$
\small H_{\text{coupling}} = J_{01}(\sigma^{(0)} \otimes \sigma^{(1)} \otimes I) + J_{12}(I \otimes \sigma^{(1)} \otimes \sigma^{(2)})
$$
Here, $J_{01}$ and $J_{12}$ represent the coupling strengths between the respective qubit pairs, and the Pauli operators ($\sigma^{(i)}$) define the nature of the interaction (e.g., $ZX$ or $YX$ coupling).

\noindent\textbf{Drive Hamiltonian~($H_{\text{drive}}(t)$):} This time-dependent term represents the malicious microwave pulses injected by the adversary (Eve) onto her assigned qubits, $q_0$ and $q_1$. The drive is modeled as:
$$
\small H_{\text{drive}}(t) = A_0 f_0(t)\, \sigma_{x}^{(0)} \otimes I \otimes I + A_1 f_1(t)\, I \otimes \sigma_{x}^{(1)} \otimes I
$$
where $A_0$ and $A_1$ are pulse amplitudes and $f_0(t)$, $f_1(t)$ are the normalized, time-varying pulse shapes controlled by Eve.

\subsection{Adversarial Strategies and Pulse Control}
The effectiveness of a crosstalk attack is highly dependent on both the timing of the malicious pulse injection and the precise shape of the pulse itself.

\noindent\textbf{Attack Timing:} The adversary can employ two primary strategies based on timing relative to the victim's operations:
\noindent \underline    {\textbf{Attacker-First:}} Eve preemptively applies malicious pulses to $q_0$ and $q_1$ \textit{before} victim; Adam, initializes or operates on $q_2$. This strategy aims to corrupt effective initial state of victim's qubit, thereby compromising all subsequent computations.

\noindent \underline    {\textbf{Victim-First:}} Eve injects her pulses \textit{after} Adam has already prepared a state on $q_2$. This approach seeks to disrupt the prepared state through crosstalk and is generally stealthier, though often less damaging.

\noindent\textbf{Pulse Control:} To execute these strategies, Eve has fine-grained control over several key pulse parameters in the rotating frame:
\begin{itemize}
    \item \textit{Pulse Amplitude ($A$)}: Amplitude of the microwave drive.
    \item \textit{Frequency Detuning ($\delta$)}: The offset between the drive frequency and the qubit's resonant frequency.
    \item \textit{Pulse Shape}: The temporal profile of the pulse. We analyze several common shapes, whose mathematical forms are detailed in Table~\ref{tab:pulse_shapes}.
    \item \textit{Shape-Specific Parameters}: Additional parameters like chirp rate ($c$), DRAG correction ($\alpha$), and pulse width ($\sigma$).
\end{itemize}

\begin{table}[!htb]
\centering
\caption{\small Mathematical definitions of pulse shapes employed in simulations.}
\vspace{-6pt}
\label{tab:pulse_shapes}
\scriptsize
\resizebox{0.48\textwidth}{!}{%
\begin{tabular}{ll}
\toprule
\textbf{Shape} & \textbf{Function $f(t)$} \\
\midrule
Cosine & $A \cos(\delta t)$ \\
Gaussian & $A \exp\left(-\frac{(t-0.5)^2}{2\sigma^2}\right)$ \\
Square & $A$ for $0.3 \le t \le 0.7$, else $0$ \\
Chirp & $A \cos\left[(\delta + c \cdot t)t\right]$ \\
DRAG & $A\left[ \exp\left(-\frac{(t-0.5)^2}{2\sigma^2}\right) - \frac{\alpha(t-0.5)}{\sigma^2}\exp\left(-\frac{(t-0.5)^2}{2\sigma^2}\right)\right]$ \\
\bottomrule
\end{tabular}%
}
\vspace{-9pt}
\end{table}

\subsection{Numerical Simulation and Channel Identification}
We simulate the system's evolution by numerically solving the time-dependent Schrödinger equation:
$$
i\hbar\, \frac{\partial}{\partial t}\ket{\psi(t)} = H(t)\ket{\psi(t)}
$$
Starting from a known initial state $\ket{\psi(0)}$, we integrate this equation to find the final state $\ket{\psi(1)}$ after the pulse sequence completes. The simulations were performed using QuTiP's \texttt{mesolve} solver, assuming purely unitary evolution (no decoherence) to isolate the effects of coherent crosstalk~\cite{li2022pulse}. The total evolution time is normalized to $t \in [0, 1]$ and discretized into 50 steps.

\section{Identifying Dominant Crosstalk Coupling}
To guide our analysis, we first performed a systematic scan to identify the most influential crosstalk channels. By applying various pulse shapes to the attacker's qubits and measuring the induced error on the victim's qubit ($q_2$), we quantified the impact using the $L_2$ norm of the difference between the ideal and attacked measurement probability distributions:
$$
\|\Delta p\|_2 = \left[ \sum_i \left( p_i^\mathrm{(attack)} - p_i^\mathrm{(ideal)} \right)^2 \right]^{1/2}
$$
where $p_i$ is the probability of measuring outcome $i$ on $q_2$.

Our preliminary results, summarized in Table~\ref{tab:coupling_scan}, show that \textbf{off-diagonal coupling terms}, particularly $Y \otimes X$ and $Z \otimes X$, produce the largest logical deviations. These channels represent the most potent vectors for adversarial influence and are therefore the primary focus of our subsequent protocol-level analysis.

\begin{table}[!htb]
\centering
\caption{\small Dominant crosstalk coupling types and their quantified influence ($\|\Delta p\|_2$) for key pulse shapes, with reference configuration: $J_{01}=J_{12}$, $q_1$ cosine drive.}
\vspace{-6pt}
\label{tab:coupling_scan}
\scriptsize
\begin{tabular}{lcc}
\toprule
Coupling Type & Pulse Shape & Influence Norm ($\|\Delta p\|_2$) \\
\midrule
Y $\otimes$ X & Chirp   & 0.0108 \\
Y $\otimes$ X & Cosine  & 0.0108 \\
Z $\otimes$ X & Chirp   & 0.0073 \\
Z $\otimes$ X & Cosine  & 0.0073 \\
Y $\otimes$ X & Square  & 0.0037 \\
Y $\otimes$ X & DRAG    & 0.0027 \\
Y $\otimes$ X & Gaussian& 0.0016 \\
Z $\otimes$ X & DRAG    & 0.0014 \\
Z $\otimes$ X & Square  & 0.0012 \\
Z $\otimes$ X & Gaussian& 0.0004 \\
\bottomrule
\end{tabular}
\vspace{-9pt}
\end{table}

\vspace{2pt}
\section{Attack Impact on Quantum Protocols}

\begin{figure}[h]
    \centering
    \scalebox{1.0}{
    \Qcircuit @C=1.0em @R=0.5em @!R {
    \nghost{q_2 : } & \lstick{q : } & \control\qw & \gate{\mathrm{V}(\lambda)} & \control\qw & \meter & \qw & \qw \\
    \nghost{} & & \raisebox{0.3em}{\text{a}} & & \raisebox{0.3em}{\text{b}} & & & \\
    \nghost{\mathrm{c} : } & \lstick{\mathrm{c} : } & \cw & \cw & \cw & \dstick{_{0}} \cw \ar @{<=} [-2,0] & \cw & \cw \\
    \nghost{} & & & & & & & \\
    }
    }
    \vspace{-2em}
    \caption{\small Experimental circuit for protocol-level attack assessment. Attack pulse is injected at either node $a$ or $b$.}
    \label{fig:exp_setup}
\end{figure}
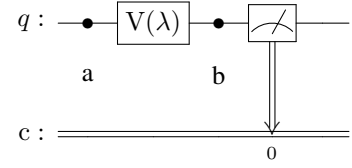

To assess the real-world impact of crosstalk, we evaluate our attack framework against two representative single-qubit quantum protocols with distinct operational characteristics: a state-preparation protocol reliant on precise analog rotations, and a classification protocol that uses large, discrete gates. In both cases, we simulate an attack where an adversarial pulse is injected either before the victim's core operation (\textbf{attacker-first}) or after it (\textbf{victim-first}), as illustrated in the general experimental setup in Figure~\ref{fig:exp_setup}. For these protocol-level simulations, we assume a moderate, symmetric coupling strength of $J_{01}=J_{12}=0.5$ and set the drive pulse detunings to zero ($\delta=0$).

\vspace{-6pt}

\subsection{Case Study 1: Biased Quantum Coin Flip}
\begin{figure}[!htb]
  \centering
  \includegraphics[width=0.95\linewidth]{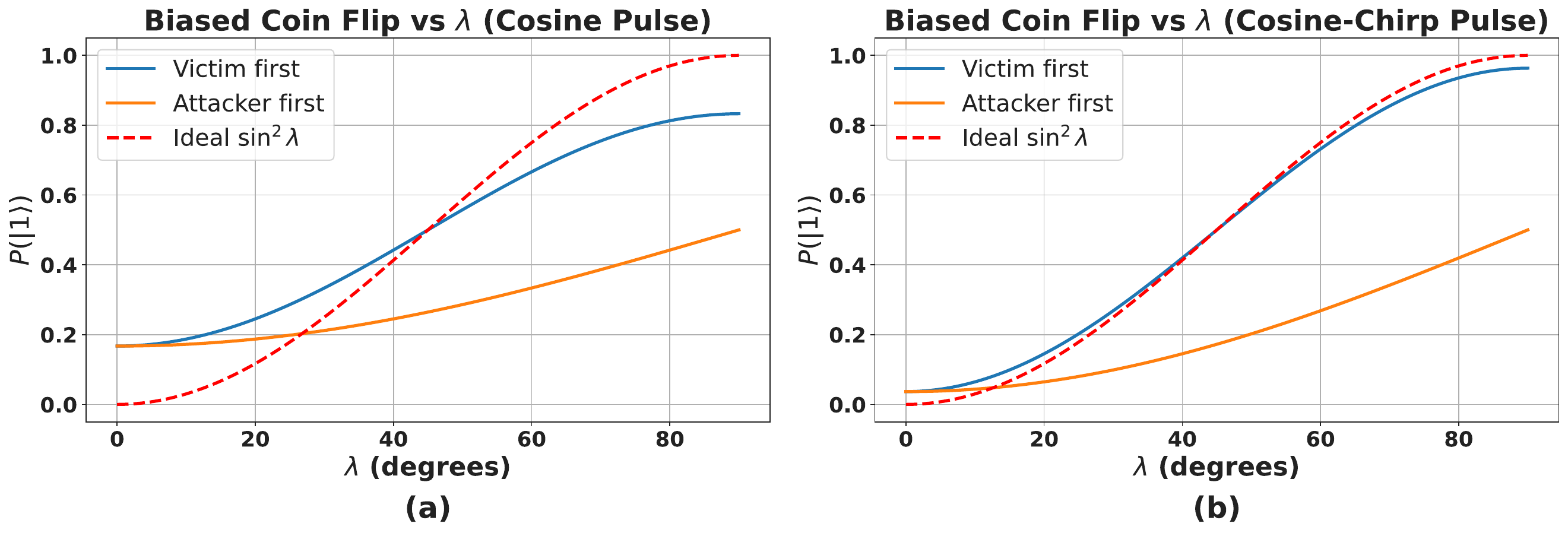}
  \vspace{-10pt}
  \caption{\small
    Crosstalk attack on the quantum coin flip protocol.
    \textbf{(a)} Moderate cosine–cosine pulses ($A_0=A_1=0.5$). 
    \textbf{(b)} Aggressive cosine–chirp pulses ($A_0=A_1=1.0$).
    Attacker-first injection causes visible bias, while victim-first remains more subtle.
  }
  \vspace{-10pt}
  \label{fig:coinflip_default}
\end{figure}

We first assess attack impact on a single-qubit state preparation protocol where the victim, Adam, prepares the state $\ket{\psi(\lambda)} = \cos\lambda\ket{0} + \sin\lambda\ket{1}$ using the rotation $V(\lambda) = e^{-i\lambda Y}$. The ideal measurement probability is $P(\ket{1}) = \sin^2\lambda$. This protocol’s reliance on the precise angle $\lambda$ makes it highly sensitive to coherent errors. \\
Fig.~\ref{fig:coinflip_default} compares two configurations: a moderate cosine–cosine pulse setup ($A_0=A_1=0.5$) and an aggressive cosine–chirp setup ($A_0=A_1=1.0$). In both cases, attacker-first injection corrupts the state before rotation, causing a measurable bias in output statistics. This is because the adversarial pulse corrupts the initial state \textit{before} the victim's rotation is applied; the victim's own gate then acts on this already-compromised state. In contrast, the victim-first strategy produces a much subtler effect, as it only slightly perturbs an already-prepared state.

These results also show that both pulse timing and shape are critical attack resources for the adversary, and  confirm that the attacker-first strategy is the most damaging vector.

\subsection{Case Study 2: XOR Quantum Classifier}
\begin{figure}[!htb]
    \centering
    \scalebox{0.8}{
    \Qcircuit @C=1.0em @R=0.2em @!R {
    \lstick{q : } & \gate{\ket{0}} & \gate{\mathrm{H}} & \gate{\mathrm{R_Z}((2x_1 - 1)\frac{\pi}{2})} & \gate{\mathrm{R_X}((2x_2 - 1)\frac{\pi}{2})} & \meter & \qw \\
    }
    }
    \caption{\small Quantum circuit for XOR classification.}
    \vspace{-6pt}
    \label{fig:xor_circuit}

\end{figure}
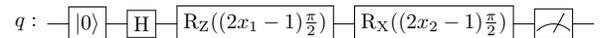

In contrast to coin flip, we also evaluate attack's impact on simple but non-trivial logical circuit: single-qubit XOR classifier~\cite{grossu2021single}. Circuit, shown in Fig.~\ref{fig:xor_circuit}, encodes two classical bits, $x_1, x_2 \in \{0, 1\}$, into angles of discrete rotations. It can be easily shown that output state is given by $|x_1 \oplus x_2\rangle$, up to some global phase. Hence, final measurement outcome deterministically classifies the XOR result. 

\begin{figure}[h!]
\centering
\includegraphics[width=0.95\linewidth]{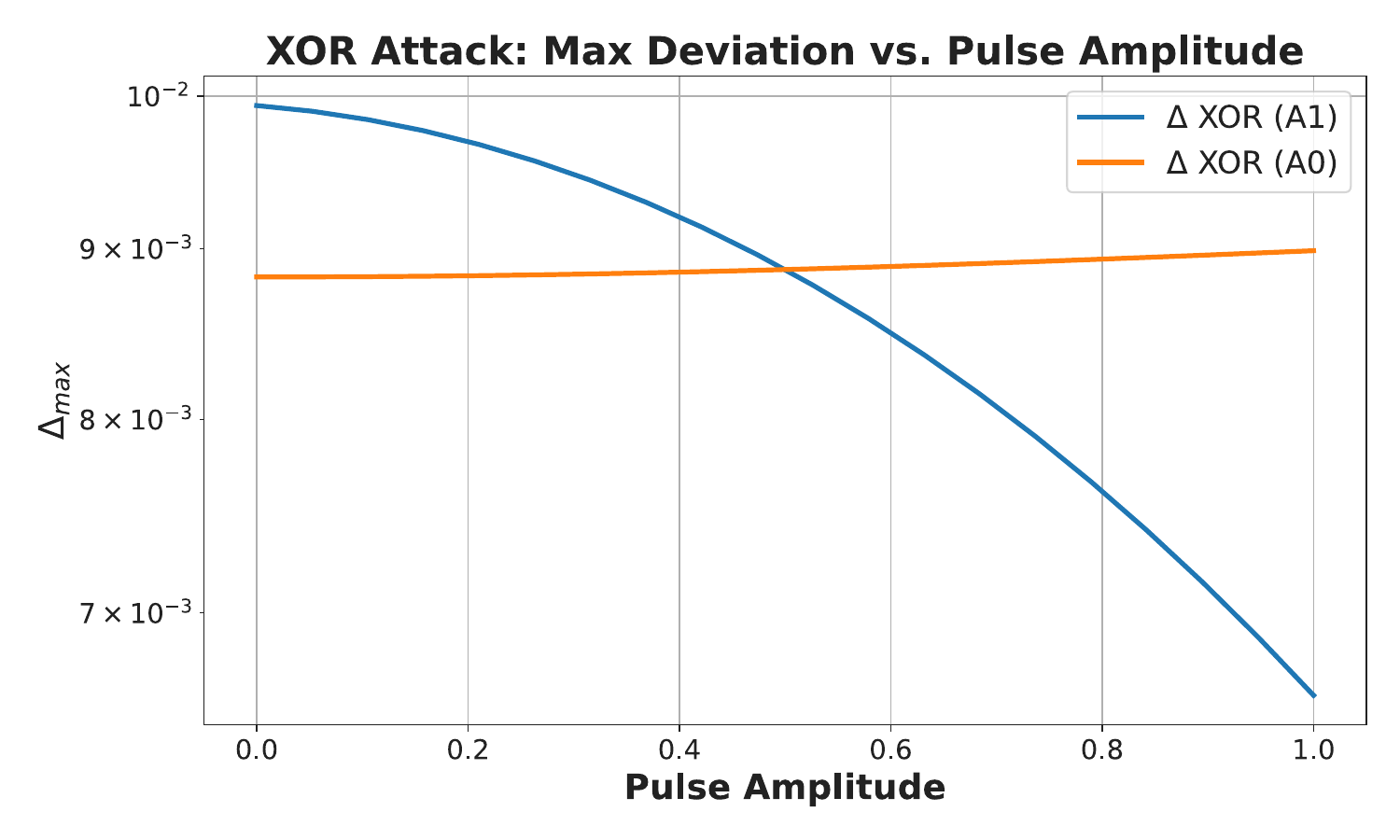}
\vspace{-12pt}
\caption{\small Maximum deviation in XOR classifier output as a function of attack amplitude observing minor changes under strong attack.}
\vspace{-8pt}
\label{fig:xor_results}
\end{figure}

\vspace{-1pt}
The XOR classifier demonstrates high resilience to the crosstalk attack. Even at maximum strength, the induced change in output statistics is minimal, on the order of $10^{-2}$, as seen in Figure~\ref{fig:xor_results}. This robustness stems from its use of discrete rotations. The small, coherent error introduced by the crosstalk pulse is insufficient to push the final state vector across the logical decision boundary that separates the computational basis states. This result shows that not all protocols are equally vulnerable and algorithmic structure plays key role in hardware-level security.
\vspace{-15pt}

\vspace{4pt}
\section{Implications \& Mitigation Strategy}

Our findings have direct implications for the security of multi-tenant quantum hardware. The core insight is that crosstalk vulnerability is protocol-dependent. The high sensitivity of the quantum coin flip protocol contrasts with the notable resilience of the XOR classifier, showing that security cannot be evaluated independently of the operations being performed at the physical level. This highlights the importance of hardware-aware security analysis for both system operators and algorithm designers.

Furthermore, our results reveal the \textit{stealthy} nature of these attacks. An adversary can strategically choose their approach to minimize the chance of detection. For instance, in the coin flip protocol (Figure~\ref{fig:coinflip_default}), an attacker using the victim-first strategy against an unbiased coin ($\lambda=45^\circ$) produces an output that is statistically identical to the ideal case, effectively masking their presence. Similarly, for the robust XOR classifier (Figure~\ref{fig:xor_results}), the impact of either attack strategy is so minimal that the resulting error would likely be indistinguishable from the device's intrinsic noise floor. This ability to either mimic ideal behavior for specific inputs or hide within system noise makes pulse-level crosstalk a particularly challenging threat to defend against.

Based on these results, we propose a practical mitigation strategy to contain crosstalk threats. First, the system can monitor for attacks by periodically executing sensitive ``canary'' circuits, such as the quantum coin flip, on idle qubits. The significant deviation from ideal statistics in the attacker-first scenario, as seen in Figures~\ref{fig:coinflip_default}, provides a clear and detectable signature of a potential attack.
Upon detecting such an anomaly, the second step is attack containment. The system should perform a high-fidelity reset operation on the affected qubit. This action effectively nullifies the adversary's attempt to corrupt the initial state and forces any subsequent malicious pulse into the much less damaging \textit{victim-first} regime. This two-step approach of detection and reset provides a simple and effective defense that contains the threat by limiting the adversary to their least effective strategy, significantly raising the difficulty of executing a successful pulse-level crosstalk attack.

\textbf{Limitations:}While our results suggest the mitigation strategy is effective under ideal conditions, the model omits decoherence and thermal noise. This abstraction isolates the coherent effects of crosstalk and highlights protocol-level vulnerabilities, but does not capture how noise might influence the attack dynamics. Future work will extend the framework to include realistic noise models and pursue experimental validation on superconducting quantum devices.

\section{Conclusion}

In this work, we have demonstrated that hardware crosstalk in multi-tenant quantum systems can be weaponized as a stealthy, pulse-level attack vector. By systematically studying coupling types, pulse parameters, and protocol-level impact, we identified which configurations have higher impact and show that protocol vulnerability can vary greatly. While protocols involving precise state preparation are most at risk, others such as XOR classification exhibit notable resilience. Our detection and reset strategy provides a simple and effective defense that can be integrated into existing quantum cloud platforms. These findings highlight the importance of hardware-aware security and suggest directions for future work on automated monitoring and hardware-level protections. This work lays the groundwork for future studies on multi-qubit protocols, realistic noise integration, and experimental attack validation on current superconducting platforms. Ultimately, securing the future of quantum computing requires a paradigm shift from viewing hardware imperfections merely as noise to recognizing them as potential security vulnerabilities.

\balance
\scriptsize

\bibliographystyle{IEEEtran}

\bibliography{references}

\end{document}